\documentclass[a4paper]{article}

\usepackage{fleqn}
\usepackage{epsfig,a4}

\def\text#1{\mbox{#1}}

\def\restylefloat#1{}
\newcommand{\ds}{\displaystyle}
\newcommand{\abc}[1]{\mbox{#1)}\quad}

\newcommand{\bm}[1]{\mbox{\boldmath $#1$}}
\newcommand{\grad}{\mathop{\rm grad}\nolimits}
\newcommand{\Div}{\mathop{\rm div}\nolimits}
\newcommand{\deriv}[2]{\mbox{$\displaystyle
\frac{\mathop{\rm d}\nolimits #1}{\mathop{\rm d}\nolimits#2}$}}

\begin{document}  \thispagestyle{plain}
\title{%
Distribution of Dark Matter around a Central Body, Pioneer Effect
and Fifth Force }

\author{        E.Schmutzer\thanks{E-mail: eschmu@aol.com}, Jena, Germany  \\
         Friedrich Schiller University }
\date{}

\maketitle \centerline{Received 2001}

\begin{abstract}
Within the framework of the Projective Unified Field Theory the
distribution of a dark matter gas around a central body is
calculated. As a result the well-known formulas of the Newtonian
gravitational interaction are altered. This dark matter effect
leads to an additional radial force (towards the center) in the
equation of motion of a test body, being used for the explanation
of the so-called ``Pioneer effect'', measured in the solar system,
but without a convincing theoretical basis up to now. Further the
relationship of the occurring new force to the so-called ``fifth
force'' is discussed.
\\[1ex]
\end{abstract}

\section{ Idea of dark matter accretion around a center \label{sec1}}

Application of the author's 5-dimensional Projective Unified Field
Theory (PUFT), being published in a series of papers (Schmutzer
1995a, 1995b, 1999, 2000a, 2001), to a closed homogeneous
isotropic cosmological model for the whole time scale (big start
to presence) led to following numerical results (Schmutzer 2000b,
2000c), where the index $p$ means present and y year:
\begin{equation}\label{S1}
  \begin{array}{lcll}\ds
\abc{a}t_{p}&=&18\cdot 10^{9}\; \text{y}\qquad & \text{(age of the
cosmos),}\bigskip \\ \ds
    \abc{b} H_{p}&=&77.6\; \frac{\mathrm{km}}{s Mpc}\qquad
    &\text{(Hubble factor),} \bigskip \\ \ds
    \abc{c}\mu _{p}&=&3.3\cdot 10^{-27}\; \mathrm{g \,cm^{-3}}
     \qquad&\text{(mass density).}
\end{array}
\end{equation}

Empirical astrophysical estimates of the visible mass density of the normal
(mostly baryonic) matter come to the maximum value
$\approx 10^{-30}\; \mathrm{g\, cm}^{-3}$. We interpreted our result with respect to this difference in the
mass densities as a hint at the existence of a dominating dark matter part
with a mass density more than two orders of magnitude greater than that of
the normal matter.

Our hypothetical model of the cosmos investigated, roughly corresponds to
the following picture of the present cosmos: \\
Our cosmos consists of a gas of dark matter particles (dm-particles)
penetrating all matter of the cosmos, particularly also the existing compact
objects (stars, nuclei of galaxies etc.) which look like buoys in the dark
matter sea. Assuming a homogeneous gas of one sort of particles, by some
hypothetical arguments we recently were led to the following numerical
values (Schmutzer 2000b):
\begin{equation}\label{S2}
\begin{array}{llll}\ds
\bigskip
 \abc{a}m_{p}&=& 2\cdot 10^{-36}\; \textrm{g}\qquad
 &\text{(rest mass of the
dm-particle),}
\\ \ds
\abc{b} T_{p}&=&1.78\;\textrm{ K}\qquad\ &\text{(kinetic temperature of the
dm-gas).}
\end{array}
\end{equation}

As a result of the attractive interaction of the dm-particles with the
compact object considered we expect a statistical distribution of the
dm-particles within and around the compact object being embedded in this
dm-gas. Let us introduce the notion ``accretion cloud'' (Umwolke) for this
part of the dm-gas which by accretion exceeds the cosmological average
distribution of the dm-gas. For simplicity we model the compact object by a
homogeneous mass sphere. Further for simplicity we apply the
Boltzmann-Maxwell statistics describing spinless particles without
degeneracy.

Let us here mention that in a different context and on a different
theoretical basis H. Dehnen et al. (Dehnen 1995) used the Bose-Einstein
statistics (preferred by physical arguments) and the Fermi-Dirac statistics
(dropped) in treating the dark matter in galaxies in order to find a
theoretical concept in understanding the problem of the ``flat rotation
curves'' of stars in galaxies. In interesting papers detailled
quantum-statistical calculations were performed. We think that for a rough
understanding of our approach it is legitimate to simplify the situation by
referring to the Boltzmann-Maxwell statistics.

\section{Basic equations}

In the following PUFT is specialized to the nonrelativistic and weak field
case, but we take into account the scalaric effects described by the
scalaric field function $\sigma $. Then the gravitational field equation is
given by
\begin{eqnarray}\label{S3}
\Delta \Phi =4\pi \gamma _{N}\mu
\end{eqnarray}
($\Phi (\bm{r},t)$ local Newtonian gravitational field of the
central body, $\mu $ mass density of the central body, $\gamma
_{N}$ Newtonian gravitational constant as a true constant).

The scalaric field may consist of two parts according to
\begin{eqnarray}\label{S4}
\abc{a}\sigma =\sigma _{c}+s \qquad\text{with} \qquad \abc{b}|s|\ll |\sigma _{c}|\; ,
\end{eqnarray}
where $\sigma _{c}(t)$ is the global scalaric cosmological field and
$s(\bm{r},t)$ the local scalaric field of the central body. Then the
scalaric field equation reads
\begin{eqnarray}\label{S5}
 \Delta \sigma =\Delta s=-\frac{4\pi \gamma _{N}\mu }{\sigma _{c}c^{2}}.
\end{eqnarray}
In the general case of a test body moving in the external fields $\Phi $ and
$s$ as well as in eventually existing external electromagnetic fields
$\bm{E}$ and $\bm{B}$ (neglecting in all fields the back reaction)
the equation of motion reads:
\begin{eqnarray}\label{S6}
M\left(\deriv{\bm{v}}{t}+\grad \Phi + \frac{c^2}{\sigma_c}\grad s
+\bm{v} \deriv{\ln \sigma _{c}}{t}\right)
=Q\left(\bm{E}+\frac{\bm{v}\times \bm{B}}{c}\right)
\end{eqnarray}
($M$ inertial mass, $Q$ electric charge, $\bm{v}$ velocity of the test
body). Let us mention that the last term on the left hand side of this
equation of motion leads to ``bremsheat production'', particularly also in
celestial bodies, investigated in detail in previous papers (Schmutzer
2000c, Schmutzer 2001).

In the following we refer to a test body without electric charge, i.e. $Q=0.$
Because of the same mathematical structure of both the equations (\ref{S3}) and
(\ref{S5}), in this approximation it is possible to introduce the combined field
function
\begin{eqnarray}\label{S7}
\Psi =\Phi \left(1-\frac{1}{\sigma _{c}^{2}}\right)+
c^{2}\ln \sigma _{c}+\frac{1}{\sigma _{c}^{2}}\Phi _{c}
\end{eqnarray}
($\Phi _{c}(t)$ cosmological value of the Newtonian potential). Hence we
obtain the common field equation
\begin{eqnarray}\label{S8}
\Delta \Psi =4\pi \gamma _{N}\left(1-\frac{1}{\sigma _{c}^{2}}\right)
\mu
\end{eqnarray}
and the equation of motion
\begin{eqnarray}\label{S9}
\deriv{\bm{v}}{t}+\grad\Psi +\bm{v}\deriv{\ln \sigma _{c}}{t}=0.
\end{eqnarray}
Let us now use following notation:
\begin{eqnarray}\label{S10}
\mu =\mu _{n}+\mu _{dm}
\end{eqnarray}
($\mu _{n}$ mass density of normal matter, $\mu _{dm}$ mass density of dark
matter). Application of this latter formula to the interior ($i$) and to the
exterior ($e$) of the spherical central body leads to the equations
\begin{eqnarray}\label{S11}
\abc{a}
\mu _{i}=\mu _{ni}+\mu _{dmi}\qquad \text{and} \qquad \abc{b}
\mu
_{e}=\mu _{ne}+\mu _{dme}\, .
\end{eqnarray}
Next we split the mass densities into both the parts coming from the global
cosmological matter (normal matter and md-matter) and from the local matter
(central body and accretion cloud):
\begin{eqnarray}\label{S12}
\abc{a} \mu _{ni}=\mu _{ng}+\mu _{ns},\qquad \abc{b} \mu_{dmi}=\mu
_{dmg}+\hat{\mu }_{dmi}
\end{eqnarray}
and
\begin{eqnarray}\label{S13}
\abc{a} \mu _{ne}=\mu _{ng}\, , \qquad \abc{b} \mu _{dme}=\mu
_{dmg}+\hat{\mu }_{dme}\, .
\end{eqnarray}
Here the indices refer to:
\begin{eqnarray*}
\begin{array}{llll}
g&\rightarrow & global\ &\text{(cosmological contribution),}\\
s&\rightarrow &sphere\ &\text{(central body contribution).}
\end{array}
\end{eqnarray*}
The roof denotes the dm- contribution induced by the central body, i.e.
surplus dark matter to be treated statistically.

For our further calculations it is convenient to introduce the abbreviations
\begin{equation}\label{S14}
\begin{array}{l}\ds
\bigskip
\abc{a} \widetilde{\mu }_{e}=\mu _{G}+\hat{\mu }_{dme} \qquad
\text{with}\qquad \abc{b} \mu _{G}=\mu _{ng}+\mu _{dmg} ,
\quad\text{i.e.}
\\ \ds
\abc{c} \mu _{i}=\mu _{G}+\mu _{ns}+\hat{\mu }_{dmi}\, .
\end{array}
\end{equation}
Let us remind that our cosmological model led us to following numerical
values of the scalaric cosmological world function:
\begin{equation}\label{S15}
\begin{array}{l}\ds
\bigskip
 \abc{a}\sigma _{c}(t\approx 1000\; \mathrm{y}) \approx  3.3 \, ,\qquad
 \abc{b}\sigma _{cp}=65.188\, , \quad\text{i.e.}
\\ \ds
\abc{c} \sigma _{c}^{2}\gg 1  \qquad \text{(for the time scale under
consideration).}
\end{array}
\end{equation}
Then the field equation (\ref{S8}) for the interior and exterior reads:
\begin{eqnarray}\label{S16}
\abc{a} \Delta \Psi _{i}=4\pi \gamma _{N}\mu _{i} ,\qquad
\abc{b} \Delta \Psi_{e}=4\pi \gamma _{N}\widetilde{\mu }_{e}.
\end{eqnarray}
Further from (\ref{S7}) results
\begin{eqnarray}\label{S17}
 \Psi =\Phi +c^{2}\ln \sigma _{c}\,.
\end{eqnarray}
Next, analogously to the above formulas for the mass density of the
dm-particles we now write down the corresponding formulas for the particle
number density $n_{dm}$ being related to the mass density as follows $(m$
mass of a dm-particle):
\begin{eqnarray}\label{S18}
\abc{a} \mu _{dm}=mn_{dm}\, ,\qquad \abc{b} \mu _{dmg}=mn_{dmg}\,
, \qquad \abc{c} \hat{\mu }_{dm}=m\hat{n}_{dm}\, .
\end{eqnarray}
We find
\begin{eqnarray}\label{S19}
n_{dm}=n_{dmg}+\hat{n}_{dm} \, .
\end{eqnarray}
In (\ref{S18}) and (\ref{S19}) we omitted the idices $i$ and $e.$

Since the time dependence of the cosmological quantities (e.g. of $\sigma
_{c}(t)$ ) is extremly slow (adiabatic time dependence), compared with the
time dependence of the cosmogonic processes of the cosmic objects, we omit
explicit writing of the time $t$ in such cosmological quantities.
Particularly constants of spatial integration implicitly involve this
adiabatic time dependence.

\section{Statistical dm-particle distribution in the accretion cloud}
\subsection{Distribution function}

According to our accretion concept presented in Section \ref{sec1}
the statistical
dm-particle number density for the case of spherical symmetry reads (k
Boltzmann constant):
\begin{eqnarray}\label{S20}
n\equiv \hat{n}_{dm}=\bar{n}\left[\exp \left(-\frac{m\chi (r)}
{\text{k}T}\right)-1\right]\, ,
\end{eqnarray}
where $\bar{n}$ denotes the dm-particle number density far away
from the cental body:
\begin{eqnarray}\label{S21}
\abc{a} \bar{n}=n(r=\infty )\qquad \text{with} \qquad \abc{b} \chi
_{\infty }\equiv\chi _{e}(r=\infty )=0\, .
\end{eqnarray}
The potental function $\chi $ used is related to $\Psi $ by
\begin{eqnarray}\label{S22}
\abc{a} \Psi =\chi +\Psi _{c}\, ,\qquad \text{where} \qquad
\abc{b} \Psi _{c}=\Phi
_{c}+c^{2}\ln \sigma _{c}\, .
\end{eqnarray}
Using (\ref{S20}) we can write the equations (\ref{S16}) as
\begin{equation}\label{S23}
\begin{array}{lll}\ds
\bigskip
\abc{a}\Delta \chi _{i}&=&4\pi \gamma _{N} \left[\mu _{0}+m\bar{n}
\left\{\exp (-\frac{m\chi _{i}}{\text{k}T})-1\right\}\right]\, ,
\\ \ds
\abc{b} \Delta \chi _{e}&=&4\pi \gamma _{N}
\left[\mu _{G}+m\bar{n}\left\{\exp (-\frac{%
m\chi _{e}}{\text{k}T})-1\right\}\right]\, ,
\end{array}
\end{equation}
where the quantity
\begin{eqnarray}\label{S24}
\mu _{0}=\mu _{G}+\mu _{ns}=const
\end{eqnarray}
(referring to a homogeneous sphere) means the mass density of the sphere.

Let us for the following use the approximation assumption
\begin{eqnarray}\label{S25}
\left|\frac{m\chi }{\text{k}T}\right|\ll 1
\end{eqnarray}
being well fulfilled in our further applications. Hence series expansion of
(\ref{S20}) leads to the linearity
\begin{eqnarray}\label{S26}
n=-\frac{\bar{n}m\chi }{\text{k}T}
\end{eqnarray}
between $n$ and $\chi $. This simplification means that we are able to
perform the further calculations analytically.
Then the equations (\ref{S23}) commonly treated (omitting the indices $i$ and $e$)
read
\begin{eqnarray}\label{S27}
 \Delta \chi +\kappa ^{2}\chi =4\pi \gamma _{N}\hat{\mu }\, ,
\end{eqnarray}
where $\hat{\mu }$ means $\mu _{0}$ for the interior or $\mu _{G}$
for the exterior, and $\kappa ^{2}$ is defined by
\begin{eqnarray}\label{S28}
 \kappa ^{2}=\frac{4\pi \gamma _{N}m^{2}\bar{n}}{\text{k}T}\,.
\end{eqnarray}

Let us here mention that the author's dissertation had referred
to the theory of strong electrolytes, where
according to the Debye-Milner theory the statistical treatment of a
(negative) ion cloud around a fixed (positive) ion (and vice versa) plays an
important role. This knowledge induced the author's
idea to treat the dark matter gas analogously. But in the first moment he
was surprised that, though between positive and negative electric charges as
well as between masses attractive forces act, in his basic equation (\ref{S27}) the
second term on the left hand side appeared with a positive sign, in contrast
to the theory of strong electrolytes. The consequence of this fact is that
in the Debye-Milner theory a Yukawa-like term with exponential decrease
occurs, whereas in this theory an interesting phenomenon of periodicity is
met.

In solving the equation (\ref{S27}) we first go over to spherical polar coordinates:
\begin{eqnarray}\label{S29}
\deriv{}{r}\left(r^{2}\deriv{\chi }{r}\right)+ \kappa^{2}r^{2}\chi
=4\pi \gamma _{N}\hat{\mu }r^{2}\,.
\end{eqnarray}

\subsection{Interior}

In this case the substitution $\hat{\mu }\rightarrow \mu _{0}$ has
to be applied. Then equation (\ref{S29}) reads
\begin{eqnarray}\label{S30}
\deriv{}{r}\left(r^{2}\deriv{\chi_i }{r}\right)+
\kappa^{2}r^{2}\chi_i =4\pi \gamma _{N}\mu r^{2}\,.
\end{eqnarray}
The solution of this inhomogenious differential equation is
\begin{eqnarray}\label{S31}
\chi _{i}=\frac{B_{1}}{r}\cos (\kappa r)+\frac{B_{2}}{r}
\sin(\kappa r)+B_{0}
\end{eqnarray}
with
\begin{eqnarray}\label{S32}
\abc{a} B_{0}=\frac{\mu _{0}\text{k}T}{m^{2}\bar{n}}=
\frac{3\gamma _{N}M_{c}}{\kappa ^{2}r_{0}^{3}}\,, \qquad
\text{where} \qquad \abc{b} M_{c}=\frac{4\pi \mu_{0}r_{0}^{3}}{3}
\end{eqnarray}
is the mass of the central body with radius $r_{0}$ ($B_{1}$and $B_{2}$
constants of integration). Regularity at $r=0$ leads to $B_{1}=0$ , i.e.
\begin{eqnarray}\label{S33}
\chi _{i}=\frac{B_{2}}{r}\sin (\kappa r)+B_{0}\, .
\end{eqnarray}
Differentiaton gives
\begin{eqnarray}\label{S34}
\deriv{\chi _{i}}{r}=\frac{B_{2}}{r^{2}}\{\kappa
r\cos (\kappa r)-\sin (\kappa r)\}\,.
\end{eqnarray}

\subsection{Exterior}

\begin{figure}[]
\centering \psfig{file=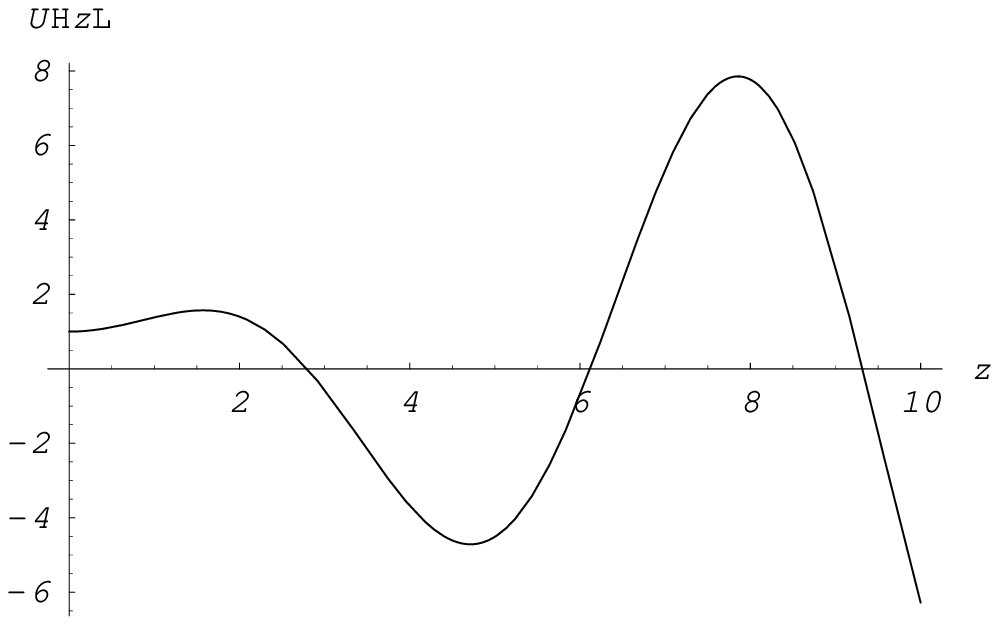,width=4in} \vspace*{0.5cm}
\caption{ Dark matter basic function U } \label{UFkt}
\end{figure}

\begin{figure}[]
\centering \psfig{file=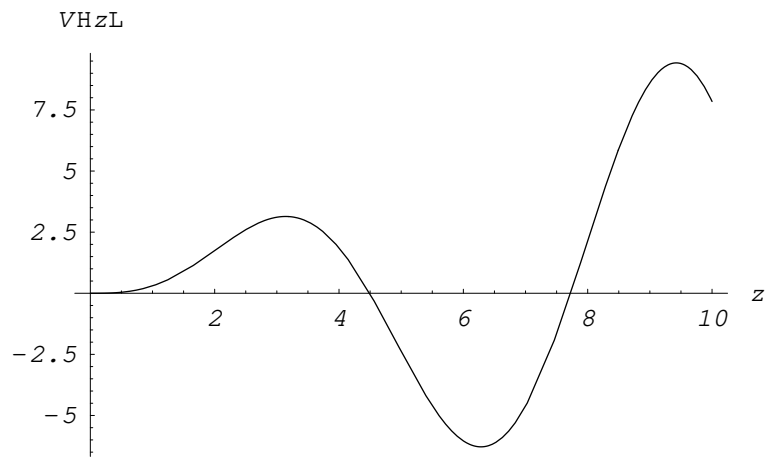,width=4in} \vspace*{0.5cm}
\caption{ Dark matter basic function V} \label{VFkt}
\end{figure}

This case differs from the previous one by the substitution
$\hat{\mu }\rightarrow \mu _{G}$. Analogously we receive the
solution
\begin{eqnarray}\label{S35}
\chi _{e}=\frac{A_{1}}{r}\cos (\kappa r)+\frac{A_{2}}{r}\sin (\kappa
r)+A_{0}
\end{eqnarray}
with
\begin{eqnarray}\label{S36}
A_{0}=\frac{\mu _{G}\text{k}T}{m^{2}\bar{n}}
\end{eqnarray}
($A_{1}$ and $A_{2}$ constants of integration). For physical reasons this
formally correct result has to be changed, because it is not in accordance
with the boundary condition (\ref{S21}b): We know that the global source $\mu _{G}$
has already be taken into account by solving the cosmological differential
equations. Therefore in this local context we have to drop the inhomogenious
contribution by setting $A_{0}=0,$ i.e.
\begin{eqnarray}\label{S37}
 \chi _{e}=\frac{A_{1}}{r}\cos (\kappa r)+\frac{A_{2}}{r}
 \sin (\kappa r)\, .
\end{eqnarray}
Differentiation leads to
\begin{eqnarray}\label{S38}
\deriv{\chi _{e}}{r}=\left(A_{2}\kappa -\frac{A_{1}}{r}\right)
\frac{\cos (\kappa r)}{r}-\left(A_{1}\kappa +\frac{A_{2}}{r}\right)
\frac{\sin (\kappa r)}{r}\, .
\end{eqnarray}
In performing these calculations we realized that following basic functions
play an important role:
\begin{eqnarray}\label{S39}
\abc{a}U(z)=z\sin z+\cos z \qquad \text{and} \qquad
\abc{b} V(z)=\sin z-z\cos z\, .
\end{eqnarray}
Figures \ref{UFkt} and \ref{VFkt} show the plots of the functions $U(z)$ and $V(z).$

\restylefloat{figure} 

\restylefloat{figure} 

\subsection{ Continuity conditions and the final form of the
potentials and  of the dm-particle number densities}

At the surface of the spherical central body following continuity conditions
have to be fulfilled:
\begin{eqnarray}\label{S40}
 \abc{a} \chi _{i}(r_{0})=\chi _{e}(r_{0})\, , \qquad
 \abc{b} \deriv{\chi_{i}(r_{0})}{r}=\deriv{\chi _{e}(r_{0})}{r}\,.
\end{eqnarray}
If these conditions are satisfied, then because of (\ref{S20}) the analogous
conditions for the dm-particle number densities result:
\begin{eqnarray}\label{S41}
\abc{a} n_{i}(r_{0})=n_{e}(r_{0})\, ,\qquad
\abc{b}\deriv{n_{i}(r_{0})}{r}=\deriv{n_{e}(r_{0})}{r}\, ,
\end{eqnarray}
i.e. thus the continuity of the dm-particle number density is guaranteed,
too. Looking at (\ref{S33}) and (\ref{S37}) we recognize that we have to cope with the 3
constants of integration $B_{2}$ , $A_{1}$ and $A_{2}.$ From both
the conditions (40) we obtain the relations
\begin{eqnarray}\label{S42}
\abc{a} A_{1}=-\frac{B_{0}}{\kappa }V(\kappa r_{0})\, ,\qquad
\abc{b}B_{2}-A_{2}=-\frac{B_{0}}{\kappa }U(\kappa r_{0})\,.
\end{eqnarray}
For physical reasons it seems to be sensible to accept that the potential in
the whole space (interior and exterior) may be determined solely by the
central mass. This means setting $A_{2}=0$.  Then finally the decisive
potentials (\ref{S33}) and (\ref{S37}) including their derivatives
(\ref{S34}) and (\ref{S38}) read:
\begin{equation}\label{S43}
\begin{array}{l}\ds
\bigskip
\abc{a}\chi _{i}=B_{0}\left\{1-\frac{1}{\kappa r}\,U(\kappa r_{0})
\sin (\kappa r)\right\}\, ,
\\ \ds
\abc{b}\deriv{\chi _{i}}{r}=\frac{B_{0}}{\kappa r^{2}}\,U(\kappa
r_{0})V(\kappa r)\, ;
\end{array}
\end{equation}
\begin{equation}\label{S44}
\begin{array}{l}\ds
\bigskip
\abc{a}  \chi _{e}=-\frac{B_{0}}{\kappa r}V(\kappa r_{0})\cos (\kappa r)\,,
\\ \ds
\abc{b}\deriv{\chi _{e}}{r}=\frac{B_{0}}{\kappa r^{2}}V(\kappa
r_{0})U(\kappa r) \, .
\end{array}
\end{equation}
Let us mention that in the center of the central body the equations
\begin{eqnarray}\label{S45}
 \abc{a} \chi _{i}(r=0)=B_{0}\{1-U(\kappa r_{0})\}\, ,\qquad
 \abc{b} \deriv{\chi _{i}(r=0)}{r}=0
\end{eqnarray}
are valid.

Furthermore, according to (\ref{S26}) we also present with the help of (\ref{S43}a)
and
(\ref{S44}a) the final form of the distribution functions $n_{i}$ and $n_{e}$:
\begin{equation}\label{S46}
\begin{array}{l}\ds
\bigskip
 \abc{a} n_{i}=-\frac{\bar{n}mB_{0}}{\text{k}T}
 \left\{1-\frac{1}{\kappa r}U(\kappa r_{0})\sin (\kappa r)\right\}\, ,
\\ \ds
\abc{b} n_{e}=\frac{\bar{n}mB_{0}}{\text{k}T\kappa r}V(\kappa
r_{0}) \cos (\kappa r)\, .
\end{array}
\end{equation}

\subsection{Series expansions}

\begin{figure}[]
\centering \psfig{file=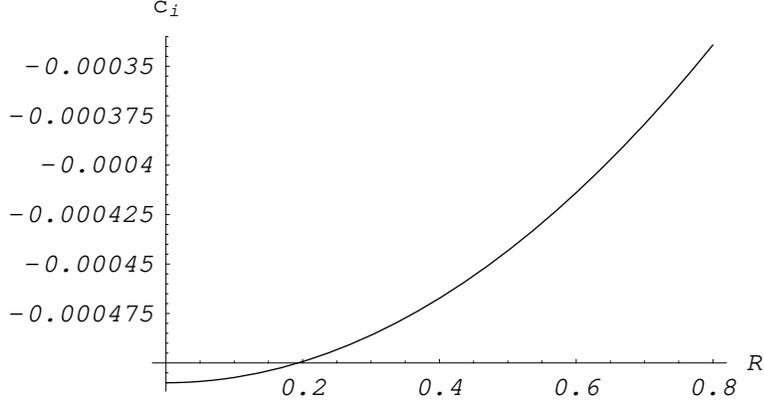,width=4in} \vspace*{0.5cm}
\caption{ Dependence of the interior potential on radius R }
\label{chi}
\end{figure}

\begin{figure}[]
\centering
\psfig{file=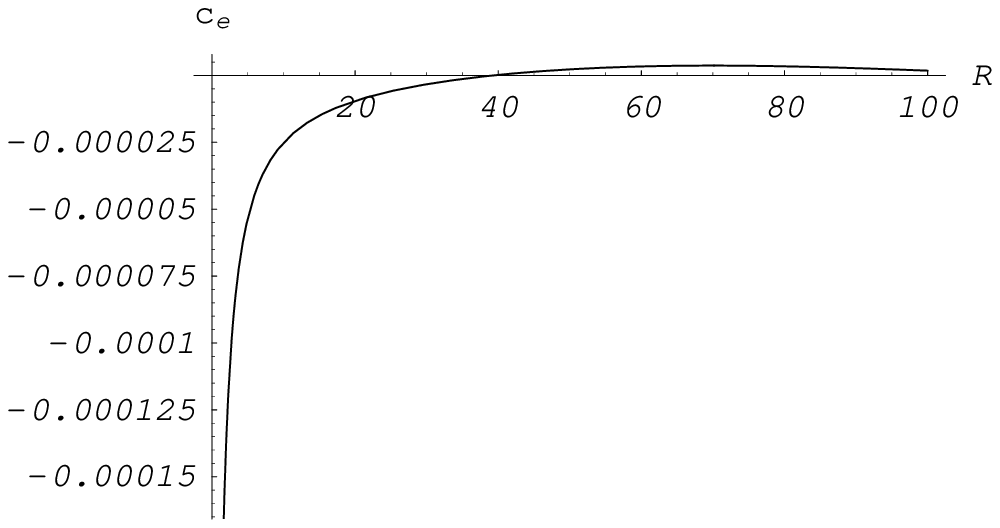,width=4in} \vspace*{0.5cm}
\caption{ Dependence of the exterior potential on radius R }
\label{che}
\end{figure}

\begin{figure}[]
\centering
\psfig{file=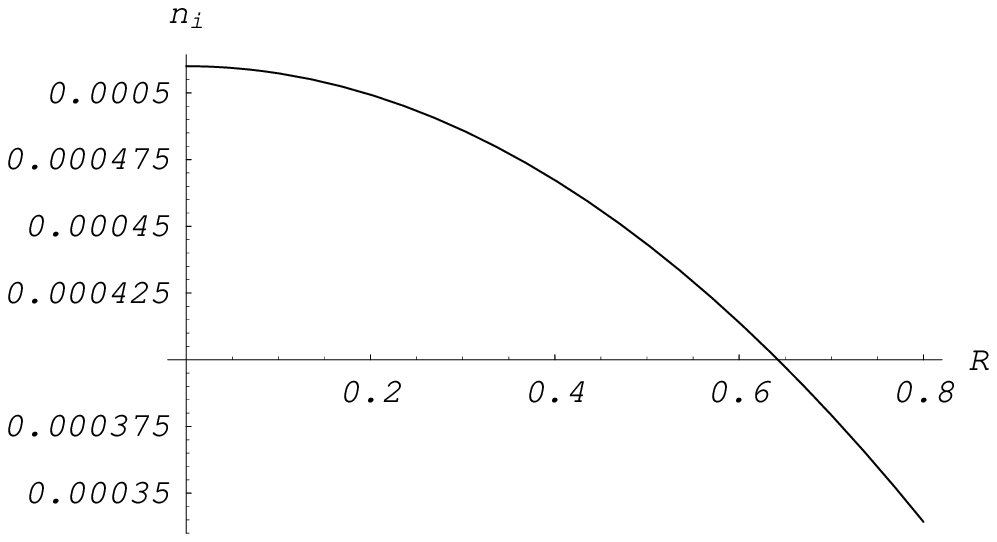,width=4in} \vspace*{0.5cm}
\caption{ Dependence of the interior dm-particle number density on radius R }
\label{ni}
\end{figure}

\begin{figure}[]
\centering
\psfig{file=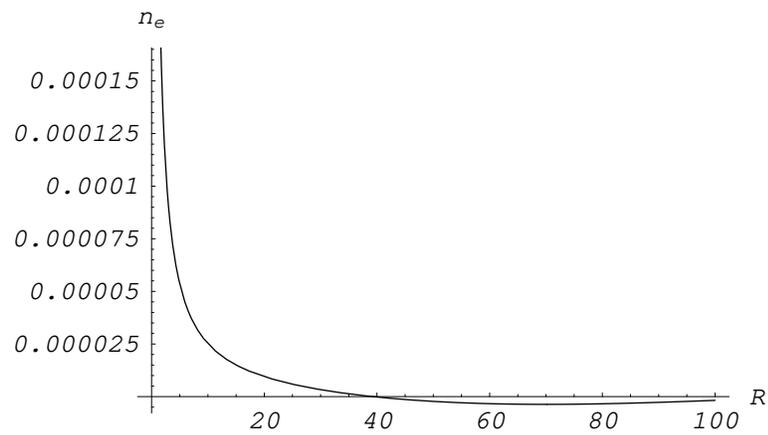,width=4in} \vspace*{0.5cm}
\caption{ Dependence of the exterior dm-particle number density on radius R }
\label{ne}
\end{figure}

For various applications it is convenient to know the first terms of the
series expansion of some functions presented above:
\begin{equation}\label{S47}
\begin{array}{l}\ds
\bigskip
\abc{a}U(z)=1+\frac{1}{2}z^{2}-\frac{1}{8}z^{4}\, ,
\\ \ds
 \abc{b} V(z)=\frac{1}{3}z^{3}(1-\frac{1}{10}z^{2})\, ;
\end{array}
\end{equation}
\begin{equation}\label{S48}
\begin{array}{l}\ds
\bigskip
\abc{a} \chi _{i}=-\frac{1}{2}B_{0}\left(\kappa r_{0}\right)^{2}
\left[1-\frac{1}{3}\left(\frac{r}{r_{0}}\right)^{2}-
\frac{1}{4}(\kappa
r_{0})^{2}\left\{1+\frac{2}{3}(\frac{r}{r_{0}})^{2}-
\frac{1}{15}\left(\frac{r}{r_{0}}\right)^{4}\right\}\right]\,,
\\ \ds
\abc{b} \deriv{\chi _{i}}{r}=\frac{1}{3}B_{0}\kappa ^{2}r
\left[1+\frac{1}{2}(\kappa r_{0})^{2}\left\{1-\frac{1}{5}
(\frac{r}{r_{0}})^{2}\right\}\right]\, ;
\end{array}
\end{equation}
\begin{equation}\label{S49}
\begin{array}{l}\ds
\bigskip
 \abc{a} \chi _{e}=-\frac{1}{3r}B_{0}\kappa ^{2}r_{0}^{3}
 \left[1-\frac{1}{2}(\kappa r)^{2}
 \left\{1+\frac{1}{5}\left(\frac{r_{0}}{r}\right)^{2}\right\}\right]\, ,
\\ \ds
\abc{b} \deriv{\chi _{e}}{r}=\frac{1}{3r^{2}}
B_{0}\kappa ^{2}r_{0}^{3}
\left[1+\frac{1}{2}(\kappa r)^{2}\left\{1-\frac{1}{5}
\left(\frac{r_{0}}{r}\right)^{2}\right\}\right]\, ;
\end{array}
\end{equation}
\begin{eqnarray}\label{S50}
\chi _{i}(r=0)=-\frac{1}{2}B_{0}(\kappa r_{0})^{2}
\left[1-\frac{1}{4}(\kappa r_{0})^{2}\right]\, .
\end{eqnarray}
One should realize that the special case $\kappa =0$ of
(\ref{S48}a) and (\ref{S49}a) corresponds to usual Newtonian
physics with the results
\begin{equation}\label{S51}
\begin{array}{lll}\ds
\bigskip
\abc{a} \chi _{Ni}&=&-\ds\frac{1}{2}B_{0}\left(\kappa
r_{0}\right)^{2}
    \left[1-\frac{1}{3}\left(\frac{r}{r_{0}}\right)^{2}\right]
    =-\frac{3\gamma _{N}M_{c}}{2r_{0}}
    \left[1-\ds\frac{1}{3}\left(\frac{r}{r_{0}}\right)^{2}\right]\, ,\\ \ds
\abc{b}
 \chi _{Ne}&=&\ds
    -\frac{1}{3r}B_{0}\kappa ^{2}r_{0}^{3}=-\frac{\gamma_{N}M_{c}}{r}\, .
\end{array}
\end{equation}
Concluding this section, we point to the Figures \ref{chi} and
\ref{che} which show plots of the radial dependence of the
potential $\chi $ on $r$ in different ranges (qualitative
presentation for $B_{0}$ $=1$). Analogously the Figures \ref{ni}
and \ref{ne} exhibit the radial dependence of the dm-particle
number density $n$ on $r$ (again qualitative presentation for
$B_{0}$ $=1$). For these Figures the value $\kappa
=0.04[(\mathrm{kpc})^{-1}]$ was chosen which apparently seems to
be appropriate for the rotation curves of the Galaxy.

\bigskip

\restylefloat{figure} 

\restylefloat{figure} 

\restylefloat{figure} 

\restylefloat{figure} 

\section{Pressure in the central body}

For physical reasons the radial course of the pressure is interesting, too.
Approaching this aim, here instead starting from the equation of motion of a
test body (\ref{S6}), we have to start from our generalization of the Navier-Stokes
equation of a continuum
\begin{equation}\label{S52}
\begin{array}{l}\ds
\bigskip
\mu \left(\deriv{\bm{v}}{t}+\grad \Phi +\frac{c^2}{\sigma_{c}}
\grad s+\bm{v}\deriv{\ln\sigma _c}{t}\right) =\rho
\left(\bm{E}+\frac{\bm{v}\times \bm{B}}{c}\right)- \grad p\, -
\\ \ds
-\eta \left(\Delta\bm{v+}\frac{1}{3}\grad \Div \bm{v}\right)
\end{array}
\end{equation}
($p$ pressure, $\rho $ electric charge density, $\eta $ viscosity).
Specialization to the case $\bm{v}=0$ (hydrostatics), $\rho =0$
(electro-neutrality), $\eta =0$ (perfect fluid) and mass homogeneity leads
to the equation
\begin{equation}\label{S53}
\grad\left(\Phi +\frac{c^2}{\sigma _c}s+\frac{p}{\mu _0}\right)=0
\end{equation}
 or
\begin{eqnarray}\label{S54}
\abc{a} \grad\left(\chi +\frac{p}{\mu _{0}}\right)=0\, ,\qquad\
    \text{i.e.} \qquad
    \abc{b} p=-\mu _{0}\chi +\bar{p}
\end{eqnarray}
($\bar{p}$ constant of integration). Using (\ref{S43}a) we find
\begin{eqnarray}\label{S55}
p=\mu _{0}B_{0}\left\{\frac{1}{\kappa r}U(\kappa r_{0}) \sin
(\kappa r)-1\right\}+\bar{p} \, .
\end{eqnarray}
The postulate of vanishing pressure at the surface of the sphere,
$p(r=r_{0})=0$, leads to the result
\begin{eqnarray}\label{S56}
p=\mu _{0}B_{0}U(\kappa r_{0})\left[\frac{\sin (\kappa r)}{\kappa r}-\frac{\sin
(\kappa r_{0})}{\kappa r_{0}}\right]
\end{eqnarray}
\bigskip
with
\begin{eqnarray}\label{S57}
p(r=0)=\mu _{0}B_{0}U(\kappa r_{0})
\left[1-\frac{\sin (\kappa r_{0})}{\kappa r_{0}}\right]\, .
\end{eqnarray}
Specialization to the Newtonian case $\kappa =0$ gives the result
\begin{eqnarray}\label{S58}
p_{N}=\frac{\gamma _{N}M_c\mu _0}{2r_0}
\left[1-(\frac{r}{r_0})^{2}\right]\, .
\end{eqnarray}

\section{Radial drift of a test body (``Pioneer effect'')}

Recent evaluation of radio metric data from Pioneer 10/11, Galileo
and Ulysses spacecraft indicates an anomalous, constant
acceleration acting on the spacecraft, being about 50 AU far from
the Sun, with a magnitude $\approx 8.5\cdot 10^{-8}
\mathrm{cm/s}^{2}$, directed towards the Sun (Anderson et al.
1998). Apparantly the explanation of this new effect seems to be
related to a correction of Newtonian mechanics by additional terms.
But also general-relativistic explanation approaches are under
discussion (Rosales and S\'{a}nchez-G\'{o}mez 2000). In the
following we try to find an explanation of this radial drift effect
by applying our theory.

The equation of motion of a test body (\ref{S9}) with the help of
(\ref{S22}a) takes the form
\begin{eqnarray}\label{S59}
\deriv{\bm{v}}{t}+\grad\chi +\bm{v}
\deriv{\ln \sigma_c}{t}=0\, .
\end{eqnarray}
From this equation we learn that for a spherically symmetric potential, as
investigated above, the appearing acceleration $A$ of the body considered
can be splitted into the Newtonian part $A_{N}$ and the scalaric part $A_{S}$
according to
\begin{eqnarray}\label{S60}
A=\deriv{\chi }{r}=A_{N}+A_{S}\, .
\end{eqnarray}
Referring to the interior and exterior of the central body we identify
\begin{eqnarray}\label{S61}
\abc{a}A_{Ni}=\frac{\gamma _{N}M_{c}r}{r_{0}^{3}}
\qquad \text{and}\qquad
\abc{b}
A_{Ne}=\frac{\gamma _{N}M_{c}}{r^{2}}
\end{eqnarray}
(internal and external Newtonian accelerations). With the help of (\ref{S43}b),
\ref{S48}b) and (\ref{S44}b), (\ref{S49}b) the additional scalaric parts take the shape
\begin{equation}\label{S62}
\begin{array}{l}\ds
\bigskip
 \abc{a} A_{Si}=\frac{\gamma _{N}M_{c}\kappa ^{2}}{(\kappa r_{0})^{3}}
 \left[\frac{3U(\kappa r_{0})V(\kappa r)}{(\kappa r)^{2}}-\kappa r\right]
 \approx\frac{\gamma _{N}M_{c}\kappa ^{2}r}{2r_{0}}
 \left[1-\frac{1}{5}\left(\frac{r}{r_{o}}\right)^{2}\right],
\\ \ds
\abc{b}A_{Se}=\frac{\gamma _{N}M_{c}}{r^{2}}
\left[\frac{3V(\kappa r_{0})U(\kappa r)}{(\kappa r_{0})^{3}}-1\right]
\approx \frac{\gamma_{N}M_{c}\kappa ^{2}}{2}
\left[1-\frac{1}{4}(\kappa r)^{2}\right]\,.
\end{array}
\end{equation}
Remembering that this series expansion is valid for $(\kappa r)^{2}$ $\ll 1$,
we apply this latter formula to the empirical facts mentioned above:
\begin{eqnarray}
\frac{\gamma _{N}M_{c}\kappa ^{2}}{2}
\left[1-\frac{1}{4}(\kappa r)^{2}\right]=8.5\cdot 10^{-8}
\mathrm{cm\,s}^{-2}\, .
\end{eqnarray}

\begin{figure}[]
\centering \psfig{file=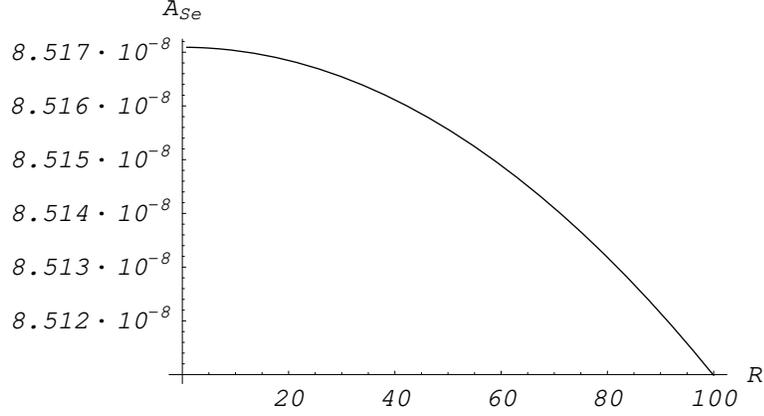,width=4in} \vspace*{0.5cm}
\caption{Dependence of the additional radial acceleration on
radius $R$ } \label{Ase}
\end{figure}

Inserting the well-known values of the Newtonian gravitational constant
$\gamma _{N}=6.68\cdot 10^{-8} \mathrm{g^{-1}\,cm^{3}s^{-2}}$, the mean distance
of the spacecraft considered from the Sun $r=50$ AU, and the mass of the Sun
$M_{c}=1.989\cdot 10^{33}\,\mathrm{g}$, we arrive at the value of $\kappa$
for the Sun (1 $\mathrm{AU = 1.496\cdot 10^{13}\, cm}$)
\begin{eqnarray}\label{S63}
\kappa =3.58\cdot 10^{-17} \,\mathrm{cm^{-1}=5.36\cdot 10^{-4}
(AU)^{-1}}\,.
\end{eqnarray}
Though our theory predicts a rather constant value of the radial
acceleration far away from the Sun, nevertheless this acceleration value
decreases at even greater distance. Fig.\ref{Ase}
shows the course of the curve in detail.
\restylefloat{figure} 

\section{Dark matter induced new interaction and the
   ``5th force'' interaction}

    Nearly twenty years ago intensive discussions on the validity of the
equivalence principle and on an eventually existing new interaction, acting
also in macrophysics, took place. Many proposals were published, where
mainly the concrete approach of Fischbach et al. met considerable attention
(Fischbach et a. 1986) who introduced the notion ``fifth force'' with the
Yukawa type potential energy between two particles $m_{1}$ and $m_{2}:$
\begin{eqnarray}\label{S64}
V(r)=-\frac{G_{\infty }m_{1}m_{2}}{r}
\left[1+\alpha \exp \left(-\frac{r}{\lambda }\right)\right]
\end{eqnarray}
($G_{\infty }$ gravitational constant at large distance, $\alpha $ and
$\lambda $ free parameters).

According to our external potential (\ref{S49}) the potential energy between such
two masses considered reads ($\gamma _{N}\rightarrow G_{\infty })$
\begin{eqnarray}\label{S65}
V_{S}=-\frac{3G_{\infty }m_{1}m_{2}}{r(\kappa r_{0})^{3}} V(\kappa
r_{0})\cos (\kappa r) \approx -\frac{G_{\infty }m_{1}m_{2}}{r}
\left[1-\frac{1}{2}(\kappa r)^{2}
\left\{1+\frac{1}{5}\left(\frac{r_{0}}{r}\right)^{2}\right\}\right]\,.
\end{eqnarray}

Since the functional structures of the potential energies
(\ref{S64}) and (\ref{S65}) are
different, the free parameters $\alpha$ , $\lambda $ and $\kappa$,
$r_{0}$ don't
correspond uniquely. Roughly speaking, only the identification
$\ds\frac{\alpha }{\lambda ^{2}}\rightarrow -\kappa ^{2}$ is possible.

\vspace{1ex}

I would like to thank Prof. Dr.A. Gorbatsievich (University of
Minsk) very much for scientific discussions and technical
help.\\[2ex]

\noindent \textbf{References}
\begin{description}
\item[]
Anderson, J.D. , Laing, Ph.A. , Lau, E.L., Liu, A.S., Nieto,
M.M.  and Turyshev, S.G.: 1998, Phys. Rev. Lett. 81, 2858
\item[] Dehnen, H., Rose, B. and Amer, K.: 1995, Astrophysics
and Space Science 234, 69
\item[]
Fischbach, E.,Sudarsky, D.,Szafer, A., Talmadge, C. and
Aronson, S.H.: 1986, Phys. Rev. Lett. 56, 3
\item[]
Rosales, J.L. and S\'{a}nchez-G\'{o}mez, J.L.: 2000 (private
communication)
\item[]
Schmutzer, E.: 1995a, Fortschritte der Physik 43, 613
\item[]
Schmutzer, E.: 1995b, Ann. Physik (Leipzig) 4, 251
\item[]
Schmutzer, E.: 1999, Astron. Nachr. 320,1
\item[]
Schmutzer, E.: 2000a, Astron. Nachr. 321,137
\item[]
Schmutzer, E.: 2000b, Astron. Nachr. 321,209
\item[]
Schmutzer, E.: 2000c, Astron. Nachr. 321,227
\item
Schmutzer, E.: 2001, General Relativity and Gravitation 33, Number
5 ( in press)
\item
\end{description}
\vspace{4ex}

\noindent Address of the author:\\[2ex] Ernst Schmutzer Cospedaer
Grund 57 \qquad D-07743 Jena Germany

\end{document}